\documentclass[conference]{IEEEtran}
\IEEEoverridecommandlockouts
\usepackage{textcomp}
\def\BibTeX{{\rm B\kern-.05em{\sc i\kern-.025em b}\kern-.08em
    T\kern-.1667em\lower.7ex\hbox{E}\kern-.125emX}}

\usepackage{amsmath,amsthm,amssymb,color,url}
\usepackage[pdftex]{graphicx}

\newtheorem{thm}{Theorem}

\def\p0{{\pmb 0}}

\long\def\comment#1{}

\newfont{\bbb}{msbm10 scaled 700}

\newfont{\bbc}{msbm10 scaled 1100}
\newcommand{\CC}{\mbox{\bbc C}}

\newcommand{\RR}{\mbox{\bbc R}}

\newcommand{\EE}{\mbox{\bbc E}}

\newcommand{\av}{{\pmb a}}

\newcommand{\cv}{{\pmb c}}

\newcommand{\ev}{{\pmb e}}
\newcommand{\fv}{{\pmb f}}

\newcommand{\hv}{{\pmb h}}

\newcommand{\kv}{{\pmb k}}

\newcommand{\qv}{{\pmb q}}

\newcommand{\sv}{{\pmb s}}
\newcommand{\tv}{{\pmb t}}
\newcommand{\uv}{{\pmb u}}
\newcommand{\vv}{{\pmb v}}
\newcommand{\wv}{{\pmb w}}

\newcommand{\yv}{{\pmb y}}
\newcommand{\zv}{{\pmb z}}
\newcommand{\zerov}{{\pmb 0}}
\newcommand{\onev}{{\pmb 1}}

\newcommand{\Am}{{\pmb A}}
\newcommand{\Bm}{{\pmb B}}

\newcommand{\Dm}{{\pmb D}}
\newcommand{\Em}{{\pmb E}}
\newcommand{\Fm}{{\pmb F}}
\newcommand{\Gm}{{\pmb G}}
\newcommand{\Hm}{{\pmb H}}
\newcommand{\Id}{{\pmb I}}

\newcommand{\Km}{{\pmb K}}

\newcommand{\Pm}{{\pmb P}}

\newcommand{\Rm}{{\pmb R}}

\newcommand{\Tm}{{\pmb T}}

\newcommand{\Vm}{{\pmb V}}

\newcommand{\Cc}{{\cal C}}

\newcommand{\Nc}{{\cal N}}

\newcommand{\etav}{\hbox{\boldmath$\eta$}}

\newcommand{\phiv}{\hbox{\boldmath$\phi$}}

\newcommand{\Gammam}{\hbox{\boldmath$\Gamma$}}
\newcommand{\Lambdam}{\hbox{\boldmath$\Lambda$}}

\newcommand{\Phim}{\hbox{\boldmath$\Phi$}}

\newcommand{\trace}{{\hbox{tr}}}

\newcommand{\herm}{{\sf H}}

\IEEEoverridecommandlockouts

\begin{document}

\title{Uplink Massive MIMO for Channels with Spatial Correlation}

\author{\IEEEauthorblockN{1\textsuperscript{st} Ansuman Adhikary}
\IEEEauthorblockA{
\textit{Qualcomm India Pvt Ltd  }\\
Hyderabad, India \\
ansumana@qti.qualcomm.com}
\and
\IEEEauthorblockN{2\textsuperscript{nd} Alexei Ashikhmin}
\IEEEauthorblockA{\textit{Bell Labs, Nokia} \\
Murray Hill, NJ \\
alexei.ashikhmin@nokia-bell-labs.com}
}

\maketitle

%

\maketitle

%

\begin{abstract}
A massive MIMO system entails a large number of base station antennas $M$ serving a much smaller number of users. This leads to large gains in spectral and energy efficiency compared with other technologies. As the number of antennas $M$ grows, the performance of such systems gets limited by pilot contamination interference \cite{marzetta2010noncooperative}. In \cite{ashikhmin2012pilot}, Large Scale Fading Precoding/Postcoding (LSFP) was proposed for mitigation of pilot contamination. In \cite{adhikary},\cite{ashikhmin} it was shown that in channels without spatial correlation (uncorrelated base station antennas) LSFP leads to large spectral-efficiency gains. In \cite{hoydis}, it was observed that if a channel has spatial correlation, then one can use this correlation to drastically reduce the pilot contamination interference in the asymptotic regime as $M\rightarrow \infty$.

In this work, we analyze the performance of Uplink (UL) transmission of massive MIMO systems with finitely many antennas $M$ for channels with spatial correlation. We extend the idea of LSFP to correlated channel models and derive  SINR expressions that depend only on slow fading channel components for such systems with and without LSFP. These simple expressions lead us to simple algorithms for transmit power optimization. As a result, we obtain a multi-fold increase in data transmission rates.
\end{abstract}

\section{Introduction}
We consider the uplink of a multicell system comprising of $L$ cells and operating in TDD (Time Division Duplexing). When the BS in each cell employs a large number of antennas, a major limiting factor in performance is pilot contamination \cite{marzetta2010noncooperative}, arising due to the users in different cells using the same pilot signals during estimation of the channel. Using LSFP \cite{ashikhmin2012pilot},\cite{adhikary},\cite{ashikhmin}, it is possible to eliminate the effects of pilot contamination (LSFP is also known as Pilot Contamination Postcoding (PCP)). LSFP requires cooperation between the cells. This cooperation is based only on slow fading channel components (like path losses) and therefore it does not depend on the number of base station antennas $M$. This can be very important in future wireless systems with very large number of antennas, especially for system operation with short wavelength signals.

Recently in \cite{hoydis}, it was shown that in channels with spatial correlation (CSC), pilot contamination is naturally mitigated, and that in the asymptotic regime, $M\rightarrow \infty$, SINRs of all users also tend to infinity, while in the case of channels without spatial correlation, SINRs tend to some finite limit \cite{marzetta2010noncooperative}.

In this work, we investigate the performance of massive MIMO systems with CSC. First, we derive SINR expressions that have simple form and depend only on slow fading channel components. Our SINR expressions can be applied to massive MIMO systems with or without LSFP. Next, we use these expressions for finding simple UL transmit power optimization algorithms. Our results show that CSC, as well as LSFP and power optimization provide a large gain in UL data transmission rates.

The rest of the paper is organized as follows. We first describe our system model in Section \ref{sec:sys-model} and in Section \ref{sec:chan_est} we remind the MMSE channel estimation via uplink pilots. Next, in Section \ref{sec:analysis-pcp} we formulate LSFP of CSC case and present our results on SINR expressions. Further, in Section \ref{sec:pcp-optz} we look at the performance improvements attained via transmit power optimization. Finally, we present our results for a realistic cellular configuration in Section \ref{sec:results} and compare uncorrelated channels and channels with spatial correlation with and without LSFP and power optimization.

\section{System Model} \label{sec:sys-model}

\begin{figure}[ht]
\centerline{\includegraphics[width=7cm]{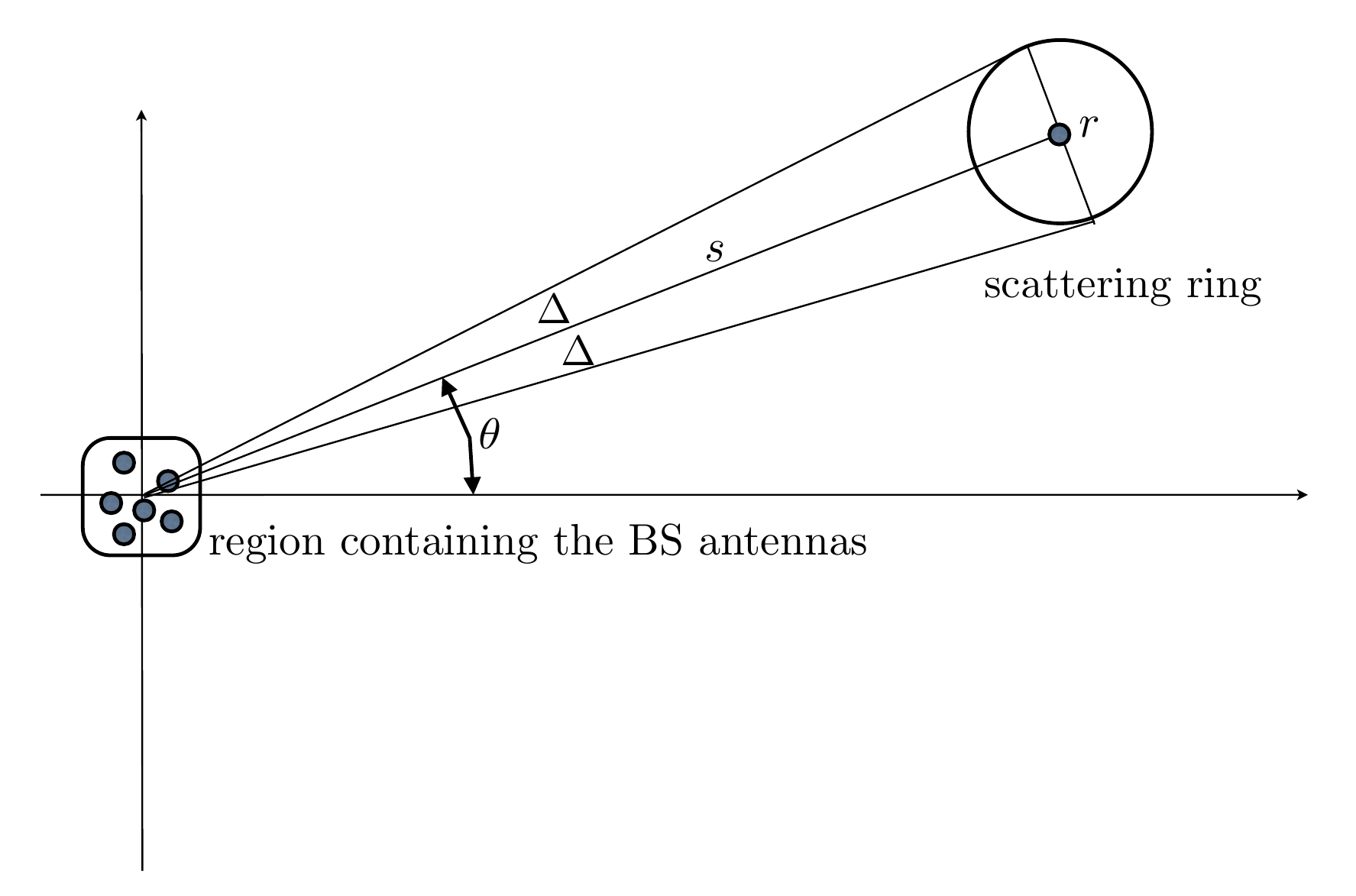}} \caption{A
user at AoA $\theta$ with a scattering ring of radius ${\sf r}$
generating a two-sided AS $\Delta$ with respect to the BS at origin.} \label{fig:one-ring-model}
\end{figure}

We assume that the network is comprised of $L$ cells and there are $K$ randomly located single antenna users in each cell. The $M \times 1$ channel vector between the $k^{\rm th}$ user in the $l^{\rm th}$ cell to the BS in the $j^{\rm th}$ cell is denoted by
 \begin{equation}
 \hv_{jkl} = \Rm_{jkl}^{\frac{1}{2}} \wv_{jkl},
 \end{equation}
 where $\Rm_{jkl} = \EE [ \hv_{jkl} \hv_{jkl}^\herm ]$ is the $M \times M$ covariance matrix (the slow fading component) and $\wv_{jkl}\sim \Cc\Nc(\zerov,\Id_M)$ denotes the fast fading component. The covariance matrix $\Rm_{jkl}$ can be decomposed as
 \begin{equation}
 \Rm_{jkl} = \beta_{jkl} \tilde{\Rm}_{jkl}
 \end{equation}
 where $\beta_{jkl}$ is the path loss,  and  $\tilde{\Rm}_{jkl}$ depends on the propagation environment between the user and BS. We model $\tilde{\Rm}_{jkl}$ according to the one ring model shown in Fig. \ref{fig:one-ring-model}, where a user located at azimuth angle $\theta_{jkl}$ and distance $\textsf{s}$ is surrounded by a ring of scatterers of radius $\textsf{r}$ such that the angular spread $\Delta = \arctan{\frac{\textsf{s}}{\textsf{r}}}$. The correlation between antennas $1 \leq m,p \leq M$ is given by \cite{shiu2000fading}
 \begin{equation} \label{eqn:covariance}
 [\Rm_{jkl}]_{m,p} = \beta_{jkl} \frac{1}{2 \Delta_{jkl}}
 \int_{-\Delta_{jkl}}^{\Delta_{jkl}} e^{j \kv^T(\alpha + \theta_{jkl}) (\uv_m - \uv_p) } d \alpha
 \end{equation}
 where $\kv(\alpha) = \frac{2 \pi}{\lambda} ( \cos(\alpha) \sin(\alpha) )^T$ is the wave vector for a planar wave impinging with AoA $\alpha$, $\lambda$ is the carrier wavelength, and $\uv_m,\uv_p \in \RR^2$ are the vectors indicating the position of BS antennas $m, p$ in the two-dimensional coordinate system.

 We assume that the BS is equipped with a uniform linear array, resulting in the covariance matrix $\Rm_{jkl}$
 \begin{equation}
 [\Rm_{jkl}]_{m,p} = \beta_{jkl} \frac{1}{2 \Delta_{jkl}} \int_{-\Delta_{jkl}}^{\Delta_{jkl}} e^{j 2 \pi D \sin(\alpha + \theta_{jkl} )(m - p) } d \alpha
 \end{equation}
 where $D$ denotes the smallest distance between the BS antennas, normalized by the carrier wavelength $\lambda$. $\beta_{jkl}$ is modeled according to the 3GPP-LTE standard for urban macro with frequency $f_c = 850 MHz$.
 \begin{equation} \label{eqn:path-loss}
 10 \log_{10} (\beta_{jkl}) = -127.8 - 35 \log_{10}(d_{jkl}) + X_{jkl}
 \end{equation}
 where $d_{jkl}$ is measured in kms and $X_{jkl} \sim \Cc \Nc (0, \sigma_{\rm shad}^2)$ represents the shadowing.

 \section{Uplink Channel Estimation}\label{sec:chan_est}
 In order to ensure reliable communication between the users and the BS, the users send pilot signals which is used by the BS to estimate the channels $\hv_{jkl}$.
 We assume that the users in all the cells use the same training codebook $\Phim = [\phiv_1 \phiv_2 \ldots \phiv_K]\in \CC^{K\times K}$ comprised  of $K$ orthonormal training vectors. The received signal at the $l^{\rm th}$ BS is
 \begin{equation}
 \tv_l = \sum_{n=1}^L \Hm_{ln} \Pm_n^{\frac{1}{2}} \Phim + \zv_l
 \end{equation}
 where $\Hm_{ln} = [\hv_{l1n} \hv_{l2n} \ldots \hv_{lKn}]$, $\Pm_n = {\rm diag}(p_{1n},p_{2n}, \ldots,$ $ p_{Kn})$ is the diagonal channel matrix of the user powers in the $n^{\rm th}$ cell, and $\zv_l \sim \Cc\Nc(\zerov,\Id_M)$ is AWGN.

 Multiplying  $\tv_l$ by $\Phim^\herm$, and taking the $k^{\rm th}$ column we get
 \begin{equation}
 \tv_{kl} = \tv_l \Phim^\herm = \sum_{n=1}^L \hv_{lkn} \sqrt{p_{kn}} + \bar{\zv}_l,~\bar{\zv}_l\sim \Cc\Nc(\zerov,\Id_M).
 \end{equation}
 The MMSE estimate $\hat{\hv}_{lkm}$ of $\hv_{lkm}$ is
 \begin{equation} \label{eqn:mmse-1}
 \hat{\hv}_{lkm} = \EE[\hv_{lkm} \tv_{kl}^\herm] \EE[\tv_{kl} \tv_{kl}^\herm]^{-1} \tv_{kl}
 =\sqrt{p_{km}} \Rm_{lkm} \Km_{kl}^{-1} \tv_{kl},
 \end{equation}
 where $\Km_{kl} = \Id_M + \sum_{n=1}^L \Rm_{lkn} p_{kn}$.
 Thus, we have
 $\hv_{lkm} = \hat{\hv}_{lkm} + \ev_{lkm}$, $\ev_{lkm}$ independent of $\hat{\hv}_{lkm}$ and $\hat{\hv}_{lkm} \sim \Cc\Nc(\zerov,\Rm_{lkm} \Km_{kl}^{-1} \Rm_{lkm} p_{km} )$, $\ev_{lkm} \sim \Cc\Nc (\zerov, \Rm_{lkm} - \Rm_{lkm} \Km_{kl}^{-1} \Rm_{lkm} p_{km})$.
   Note also that $\hat{\hv}_{lkm}$ and $\hat{\hv}_{lkl}$ are correlated, with
 \begin{equation}
 \EE [\hat{\hv}_{lkm} \hat{\hv}_{lkl}^\herm] = \Rm_{lkm} \Km_{kl}^{-1} \Rm_{lkl} \sqrt{ p_{kl} p_{km} }.
 \end{equation}

 \section{Large Scale Fading Postcoding}\label{sec:analysis-pcp}

  LSFP is a way of organizing cooperation between base stations so that this cooperation is based only on slow fading components. So the traffic needed for this cooperation is independent of $M$ and OFDM tone index. The fact that slow fading components change about 40 times slower than fast fading components also reduces the needed communication traffic.

 As it will be clear from the description presented below,  we formulate all our results for LSFP with generic $L^2 \times L$  LSFP matrices $\Am_k = [\av_{k1} \av_{k2} \ldots \av_{kL}],~k=1,\ldots,K$. By $a_{kljp}$ we denote the element on the intersection of the $l$-th column and $(p-1)L + j$-th row of $\Am_k$.
 LSFP matrices with $a_{kljp} = 1$ when $j = p = l$ and $a_{kljp} = 0$ for all other indices mean that there is no cooperation between base stations, that is, we do not use LSFP.

 After transmitting pilots, all users transmit uplink data and the $l^{\rm th}$ BS receives the vector
 \begin{equation}\label{eqn:vy_l}
 \yv_l = \sum_{n=1}^L \sum_{m=1}^K \hv_{lmn} \sqrt{q_{mn}} s_{mn} + \zv_l
 \end{equation}
 where $q_{mn}$ is the power of the $m^{\rm th}$ user in the $n^{\rm th}$ cell and $s_{mn}$ is the corresponding data symbol.
 Next the $l^{\rm th}$ BS applies an $M$-dimensional receiver to $\yv_l$. In this work we assume that either Matched Filtering (MF) or Zero-Forcing (ZF) receivers are used. (The important case of MMSE receiver will be consider in a future work.)
 As a result, the $l^{\rm th}$ BS gets the estimate $\tilde{s}_{klp}$ of signals $s_{kp}$. In particular, in the case of MF receiver,
 \begin{equation}\label{eqn:tilde_s_MF}
 \tilde{s}_{klp} = \hat{\hv}_{lkp}^\herm \yv_l,
 \end{equation}
 and in the case of ZF receiver
 \begin{equation}\label{eqn:tilde_s_ZF}
 \tilde{s}_{klp} = \vv_{lkp}^\herm \yv_l,
 \end{equation}
 were $\vv_{lkp}$ denotes the $((k-1)L + p)^{\rm th}$ column of
 $$\Vm_l=\hat{\Hm}_l(\hat{\Hm}_l^\herm \hat{\Hm}_l)^{-1}, \mbox{ where }
 \hat{\Hm}_l = [\hat{\hv}_{l11} \ldots \hat{\hv}_{l1L} \hat{\hv}_{l21} \ldots \hat{\hv}_{lKL}].
 $$

Next the $l^{\rm th}$ BS sends the quantities $\tilde{s}_{klp}$ and $\EE[\hat{\hv}_{lkp}^\herm \hat{\hv}_{lkn}]$, $l,p,n=1,\ldots,L$, and $k=1,\ldots,K$, to a central controller (SC). SC forms the $L^2 \times 1$ vector $\tilde{\sv}_k = [\tilde{s}_{k11}, \ldots ,\tilde{s}_{k1L},$ $\tilde{s}_{k21}, \ldots ,\tilde{s}_{kLL}]^T$, and computes estimates
$$\hat{\sv}_k = [\hat{s}_{k1} \ldots \hat{s}_{kL}]^T =\Am_k^\herm \tilde{\sv}_k,$$
of data symbols sent by the $k^{\rm th}$ user in all cells.

 Let ${\rm SINR}_{kl}$ be the SINR of the $k^{\rm th}$ user in the $l^{\rm th}$ cell.
 Our goal is to derive estimates for ${\rm SINR}_{kl}$ with different receivers as functions of only the slow fading components. Such estimates are important for several reasons. They give an insight into the system performance, explicitly showing main sources of interference and further allowing to find bottle-necks that prevents us from further performance improvement. Next, they allow simple simulations of systems with large $M$, since we do not have to simulate $M$-dimensional receivers, but simply generate slow fading components and substitute them into the estimates of ${\rm SINR}_{kl}$. Finally, and perhaps most importantly, such estimates allow us to use power optimization algorithms that depend only on slow fading components. Typically, such algorithms are simple and they allow updating power with much less frequency than algorithms based on fast fading components.

 Let $\hat{\Am}_k=[\hat{\av}_{k1} \hat{\av}_{k2} \ldots \hat{\av}_{kL}]$ be the matrix with entries  $\hat{a}_{kljp} = a_{kljp} \sqrt{p_{kp}}$.
 We formulate our first result  without further detail due to page limit.
 \begin{thm}\label{thm:SINR_MF} If  $M$-dimensional MF receiver is used then
 \begin{equation}
{\rm SINR}_{kl}
 = \frac{\left|\sum_{j=1}^L \sum_{p=1}^L \hat{a}_{kljp}^* \trace{\left(\Rm_{jkl} \Km_{kj}^{-1} \Rm_{jkp}\right)}\right|^2 \times p_{kl} q_{kl}}
 {I_1+I_2+I_3},
 \end{equation}
 where
 \begin{align*}
 I_1=&\sum_{n=1,n\neq l}^L \left|\sum_{j=1}^L \sum_{p=1}^L \hat{a}_{kljp}^* \trace{\left(\Rm_{jkn} \Km_{kj}^{-1} \Rm_{jkp}\right)}\right|^2 p_{kn} q_{kn},\\
 I_2=&\sum_{m=1}^K \sum_{n=1}^L q_{mn} \sum_{j=1}^L \sum_{p=1}^L \sum_{p'=1}^L \hat{a}_{kljp}^* \hat{a}_{kljp'} \\
 &\cdot \trace{\left(  \Rm_{jmn} \Rm_{jkp'} \Km_{kj}^{-1} \Rm_{jkp} \right)},   \\
I_3=&\sum_{j=1}^L \sum_{p=1}^L \sum_{p'=1}^L \hat{a}_{kljp}^* \hat{a}_{kljp'} \trace{\left(\Rm_{jkp'} \Km_{kj}^{-1} \Rm_{jkp} \right)}.
\end{align*}
 \end{thm}

 Let us consider now the case of $M$-dimensional ZF receiver. After substitution of $\yv_l$ from (\ref{eqn:vy_l}) into (\ref{eqn:tilde_s_ZF}) and some computations, we obtain
 \begin{align}
\hat{s}_{kl} =& \av_{kl}^\herm \tilde{\sv}_k = \sum_{j=1}^L \sum_{p=1}^L a_{kljp}^* \tilde{s}_{kjp} \nonumber\\
=& \underbrace{\sum_{j=1}^L a_{kljl}^* \sqrt{q_{kl}} s_{kl}}_{\rm Useful \ \ Signal} + \underbrace{\sum_{n=1,n \neq l}^L \sum_{j=1}^L a_{kljn}^* \sqrt{q_{kn}} s_{kn}}_{\rm Pilot \ \ Contamination}\nonumber\\
& + \sum_{n=1}^L \sum_{m = 1}^K \sum_{j=1}^L \sum_{p=1}^L a_{kljp}^* \vv_{jkp}^\herm \ev_{jmn} \sqrt{q_{mn}} s_{mn}\\
 & \underbrace{\phantom{..................................} + \sum_{j=1}^L \sum_{p=1}^L a_{kljp}^* \vv_{jkp}^\herm \zv_{j}.}_{\rm Interference \ \ plus \ \ Noise \ \ Terms}
\end{align}
Let $T_0,T_1$, and $T_2$ be the terms in the above expression. Then the data transmission rate for the $k^{\rm th}$ user in $l^{\rm th}$ cell is
$$
R_{kl} = \EE_{\Vm_l,l \in \{1,\ldots,L\}} [ \log_2 ( 1 + \frac{\EE[|T_0|^2 | \vv_{jkp},\forall j,p\} ]}
{\EE[|T_1|^2+|T_2|^2|  | \vv_{jkp}, \forall j,p\} ]}) ].
$$
Computing expectations in the above expression and using Jensen's inequality, we obtain
\begin{align*}
R_{kl} &= \EE_{\Vm_l,l \in \{1,\ldots,L\}} [\log_2(1+{|\sum_{j=1}^L a_{kljl}^* |^2 q_{kl}\over I_1+I_2})]\\
&\ge \log_2(1+{|\sum_{j=1}^L a_{kljl}^* |^2 q_{kl}\over I_1+\EE_{\vv_{jkp},\vv_{jkp'}}[I_2]}),
\end{align*}
where
\begin{align}
I_1=&\sum_{n=1,n\neq l}^L \left|\sum_{j=1}^L a_{kljn}^* \right|^2 q_{kn},\nonumber\\
I_2=&\sum_{j=1}^L \sum_{p=1}^L \sum_{p'=1}^L a_{kljp} a_{kljp'}^* \vv_{jkp}^\herm \left[ \sum_{n=1}^L \sum_{m = 1}^K  \left( \Rm_{jmn} - \right. \right. \nonumber\\
&\left. \left. \Rm_{jmn} \Km_{mj}^{-1} \Rm_{jmn} p_{mn} \right) q_{mn} + \Id_M \right] \vv_{jkp'} \nonumber\\
=& \sum_{j=1}^L \sum_{p=1}^L \sum_{p'=1}^L a_{kljp} a_{kljp'}^* \bar{I}_2 \label{eqn:approx-val}
\end{align}
Using an approximation via random matrix theory, we obtain the following result.
\begin{thm}\label{thm:SINR_ZF}  If ZF receiver is used then
\begin{equation}
{\rm SINR}_{kl} \stackrel{M \rightarrow \infty}{=} {\rm SINR}_{kl}^{\rm approx}
\frac{\left|\sum_{j=1}^L a_{kljl}^* \right|^2 q_{kl} }{I_1+I_2},
\end{equation}
where
\begin{align}
I_1=&\sum_{n=1,n\neq l}^L \left|\sum_{j=1}^L a_{kljn}^* \right|^2 q_{kn}\nonumber\\
 I_2=&\sum_{j=1}^L \sum_{p=1}^L \sum_{p'=1}^L a_{kljp}^* a_{kljp'} \ev_p^T \Gammam_{jk} \ev_{p'}, \label{eqn:approx-val3}
\end{align}
where $\ev_p$ is the $p^{\rm th}$ column of the identity matrix $\Id_L$, and
$\Gammam_{jk}$ is provided at the very end of the Appendix, which is a function of the covariance matrices $\Rm_{jkl}$.
\end{thm}
A proof of this theorem is quite technical. We present a sketch of it in Appendix.

Theorems \ref{thm:SINR_MF} and \ref{thm:SINR_ZF} give simple expressions for SINRs, which further allow us to find optimal LSFP matrices $\Am_k$.

Let us define $\cv_{kn} = [c_{kn11} \ldots c_{kn1L} c_{kn21} \ldots c_{knLL}]$ with $c_{knjp} = \trace{\left(\Rm_{jkn} \Km_{kj}^{-1} \Rm_{jkp}\right)}$ and block diagonal matrices $\Dm_{k} = {\rm diag}(\bar{\Dm}_{1k},\bar{\Dm}_{2k},\ldots,\bar{\Dm}_{Lk})$ with
\begin{align*}
&[\bar{\Dm}_{jk}]_{p,p'}\\
=& \trace{(\Rm_{jkp'} \Km_{kj}^{-1} \Rm_{jkp} + \sum_{m=1}^K \sum_{n=1}^L \Rm_{jmn} \Rm_{jkp'} \Km_{kj}^{-1} \Rm_{jkp} q_{mn} )}.
\end{align*}
 Let us further define $\etav_{l}$ to be the $l^{\rm th}$ column of the matrix $\onev_L \otimes \Id_L$, and
block diagonal matrices $\Em_{k} = {\rm diag}(\bar{\Em}_{1k},\bar{\Em}_{2k},\ldots,\bar{\Em}_{Lk})$ with
$$[\bar{\Em}_{jk}]_{p,p'} = \ev_p^T \Gammam_{jk} \ev_{p'}.$$
Note that the matrices $\Dm_k, \Em_k$ are $L^2 \times L^2$, while the matrices $\bar{\Dm}_{lk}, \bar{\Em}_{lk}$ are $L \times L$.
\begin{thm}\label{thm:Optimal PCP MF and ZF}
For MF receiver the optimal LSFP matrices are defined by vectors
$$\hat{\av}_{kl} = ( \sum_{n=1,n \neq l}^L \cv_{kn} \cv_{kn}^\herm p_{kn} q_{kn} + \Dm_k )^{-1} \cv_{kl},$$
leading to
\begin{equation} \label{eqn:SINR}
{\rm SINR}_{kl}^{(MF)} = \cv_{kl}^\herm ( \sum_{n=1,n \neq l}^L \cv_{kn} \cv_{kn}^\herm p_{kn} q_{kn} + \Dm_k )^{-1} \cv_{kl} \cdot p_{kl} q_{kl}.
\end{equation}
For ZF receiver we have
$$\hat{\av}_{kl} = ( \sum_{n=1,n \neq l}^L \etav_{n} \etav_{n}^\herm p_{kn} q_{kn} + \Em_k )^{-1} \etav_{l},$$
giving
\begin{equation} \label{eqn:SINR-zf}
{\rm SINR}_{kl}^{(ZF)} = \etav_{l}^\herm ( \sum_{n=1,n \neq l}^L \etav_{n} \etav_{n}^\herm p_{kn} q_{kn} + \Em_k )^{-1} \etav_{l} \cdot p_{kl} q_{kl}.
\end{equation}
\end{thm}

\section{Transmit Power Optimization}\label{sec:pcp-optz}

SINR expressions presented in Theorem \ref{thm:Optimal PCP MF and ZF} allow us to find optimal transmit powers.

Due to space limit, we formulate results only for MF-receiver. Results for ZF-receiver are similar.
We consider the following optimization problem

\begin{align} \label{eqn:optz-q}
&\max_{\qv} \min_{k,l}  ~{\rm SINR}_{kl} \nonumber\\
=&\max_{\qv} \min_{k,l}  ~\cv_{kl}^\herm ( \sum_{n=1,n \neq l}^L \cv_{kn} \cv_{kn}^\herm p_{kn} q_{kn} + \Dm_k )^{-1} \cv_{kl} p_{kl} q_{kl}, \nonumber\\
&{\rm subject\ to}  ~\zerov \leq \qv \leq Q_{\max} \onev,
\end{align}
where $\qv$ is the $KL \times 1$ vector of  the user powers, and  $\onev$ is a $KL \times 1$ vector of all ones.
This optimization problem can be equivalently formulated as
\begin{align}
\max_{\qv} & ~\gamma \nonumber\\
{\rm subject\ to} & \nonumber\\
 \zerov \leq \qv& \leq Q_{\max} \onev, \label{eqn:optz-q-2_1}\\
\cv_{kl}^\herm ( \sum_{n=1\atop n \neq l}^L \cv_{kn} \cv_{kn}^\herm p_{kn} q_{kn} + \Dm_k )^{-1} \cv_{kl} p_{kl} q_{kl} &\geq \gamma,~\forall k,l.
\label{eqn:optz-q-2_2}
\end{align}
This optimization problem can be solved with the following iterative bisection algorithm:
\begin{enumerate}
	\item Set $\gamma_{\max} = \max_{k,l} || \cv_{kl} ||^2 P_{\max} Q_{\max}$ and $\gamma_{\min} = 0$.
	\item Set $\gamma = (\gamma_{\max} + \gamma_{\min})/2$.
	\item Check the feasibility of constraints (\ref{eqn:optz-q-2_1}) and (\ref{eqn:optz-q-2_2}).
	\item  If $\gamma$ is feasible, assign $\gamma_{\min} = \gamma$ and go to Step 6, else go to Step 5.
	\item Set $\gamma_{max}=\gamma$.
	\item If $\gamma_{\max} - \gamma_{\min} < \epsilon$ ($\epsilon$ is a small number), stop and output $\gamma_{\min}$.	
\end{enumerate}

For checking feasibility at Step 3 of the above algorithm, we can use the following distributed power optimization algorithm. This algorithm can be also used on its own to achieve a desired SINR target for all users, and in fact it leads to better 5~$\%$ outage rates (see Section \ref{sec:results}). The distributed algorithm is as follows:
\begin{enumerate}
	\item Set $\qv = \qv^{(0)}$ and compute ${\rm SINR}_{kl}^{(0)}$ according to
	(\ref{eqn:SINR}).
	\item At iteration $n$ compute
	$q_{kl}^{(n)} = \min\{  Q_{\rm max},$
$q_{kl}^{(n-1)}\gamma/{\rm SINR}_{kl}^{(n-1)} \}$.
	\item If $|| \qv^{(n)} - \qv^{(n-1)} ||_2 < \epsilon || \qv^{(n)} ||_2,~ \forall k,l$ stop, else go to Step 2.
\end{enumerate}
\begin{thm} The distributed algorithm always converges and if $\gamma$ is feasible, it converges to powers $q_{kl}$ that minimize total power $\sum_{k}\sum_{l} q_{kl}$.
\end{thm}

\section{Numerical Results} \label{sec:results}
We consider a cellular layout consisting of $L = 7$ cells, with $M = 100$ and $K = 5$ users.
(We are currently working on results for large networks with $L\ge 19$ and hope to present them in the final version of this conference paper.)
Each cell has a cell radius $R_c = 1$ km, with users generated randomly within the cellular coverage area. The user position determines the distance, angle of arrival and angular spread to all the base stations. The scattering radius is fixed at $r = 20$ m, and the covariance matrices are generated using (\ref{eqn:covariance}). The path loss coefficients $\beta$'s are generated according to (\ref{eqn:path-loss}), and the variance of the log normal shadowing coefficient is taken to be $\sigma_{\rm shad} = 8$ dB. The maximum transmit power of a user is taken to be $Q_{\rm max} = 200$ mW. The noise variance is given as
\begin{equation}
{\rm Noise\ Var.\ (dBm)} = -174 + 10 \log_{10} B + {\rm NF} + 2,
\end{equation}
where the bandwidth $B = 20$ MHz and ${\rm NF}=4$  is the noise figure at the BS. Based on these parameters, the SNR at the cell edge (neglecting the shadowing) is approximately $-6$ dB (taking into account a $2$ dB antenna gain). The SINR expressions for LSFP with matched filtering MF and ZF receiver are computed according to (\ref{eqn:SINR}) and (\ref{eqn:SINR-zf}) respectively. For SINR expressions without LSFP, we use (\ref{eqn:SINR}) and (\ref{eqn:SINR-zf}) with $a_{kl}$ such that $a_{kljp} = 1$ when $j = p = l$. The pilot powers and transmission powers are equal to $Q_{\max}$ for all users.

\begin{figure}
  \centering
 \includegraphics[width=7cm]{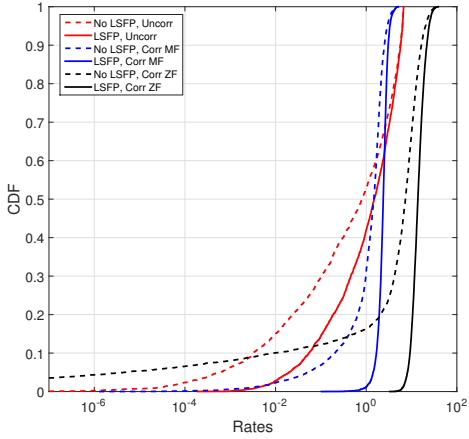}
  \caption{CDF of user rates for different scenarios}\label{fig:rates-cdf-comp}
\end{figure}

Figure \ref{fig:rates-cdf-comp} shows a comparison between the CDF of the user rates with and without LSFP between correlated and uncorrelated channels. By uncorrelated channels, we mean the covariance matrices are given as $\Rm_{jkl} = \beta_{jkl} \Id_M, \ \forall \ j,k,l$. The ``dashed'' curves correspond to the user rates without LSFP and the ``solid'' curves denote the user rates with LSFP. It can be seen from Fig. \ref{fig:rates-cdf-comp} that LSFP gives a significant improvement in the user rates compared to the scenario without LSFP. Also, user rates in the case of correlated channels are better than the uncorrelated case.

\begin{figure}
  \centering
  \includegraphics[width=7cm]{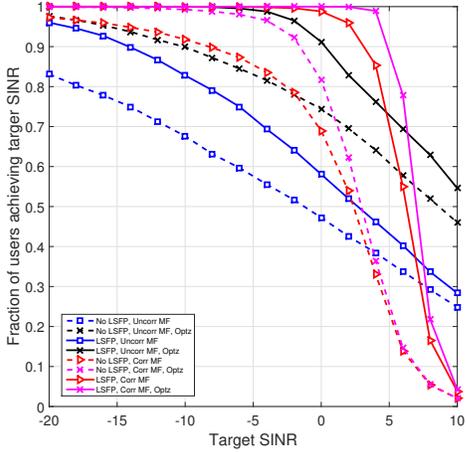}\\
  \caption{Fraction of Users achieving Target SINR for various schemes in Correlated Channels with Matched Filtering receiver. The horizontal dashed line in black denotes $5 \%$ outage.}\label{fig:pcp-dist-mf}
\end{figure}

Figure \ref{fig:pcp-dist-mf} shows the fraction of users achieving a desired SINR target for transmission schemes with MF receiver. The ``dashed'' and ``solid'' curves denote the results for without and with LSFP respectively. For obtaining the ``black'' and ``magenta'' curves, we solve the power optimization problem defined by (\ref{eqn:optz-q-2_1}) and (\ref{eqn:optz-q-2_2}) by fixing a target SINR, for uncorrelated and correlated channels respectively.

One can see that using proper power allocation schemes for MF in addition to LSFP enables increases outage SINR for both correlated and uncorrelated channels. For example, the $5 \%$ outage SINR for correlated channels without LSFP is -15 dB, whereas using LSFP, it can be increased to 2 dB, and further  to 4 dB using proper power allocation. This translates to a 40 times increase in the data rates.

\begin{figure}
  \centering
  \includegraphics[width=7cm]{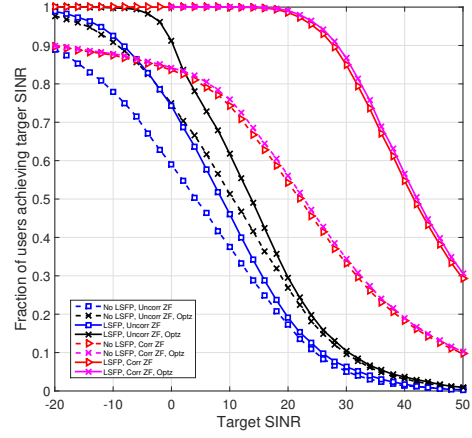}\\
  \caption{Fraction of Users achieving Target SINR for various schemes in Correlated Channels with Zero Forcing receiver. The horizontal dashed line in black denotes $5 \%$ outage.}\label{fig:pcp-dist-zf}
\end{figure}

Figure \ref{fig:pcp-dist-zf} shows results for ZF receiver. Power optimization over LSFP does not yield significant gains in the outage SINR  compared to LSFP without power optimization when channels are correlated.  However, the $5 \%$ outage SINR for LSFP with ZF gives a 21 dB improvement over the MF scenario, corresponding to a 4.5 times increase in the data rates over LSFP with MF and a 185 times increase over no LSFP with MF.

\section{Appendix}
\label{sec:det-eq-corr}

In this section, we provide an approximation to the quantity $\bar{I}_2=\vv_{jkp}^\herm \Lambdam_j \vv_{jkp'}$, $\Lambdam_j = \sum_{n=1}^L \sum_{m = 1}^K  \left( \Rm_{jmn} - \Rm_{jmn} \Km_{mj}^{-1} \Rm_{jmn} p_{mn} \right) q_{mn} + \Id_M$,  defined in
(\ref{eqn:approx-val}), in the regime $M,K\rightarrow \infty$, $M/K=\mbox{const}$. We remind that $\vv_{jkp}$ denotes the $((k-1)L + p)^{\rm th}$ column of $\Vm_j$, where
$$
\Vm_j = \hat{\Hm}_j (\hat{\Hm}_j^\herm \hat{\Hm}_j)^{-1} = \lim_{z \rightarrow 0} ( \hat{\Hm}_j \hat{\Hm}_j^\herm - M z \Id_M )^{-1} \hat{\Hm}_j.
$$
Hence, our goal is equivalent to evaluation of
\begin{align*} \label{eqn:zf-qoi}
\lim_{z \rightarrow 0} & ~\hat{\hv}_{jkp}^\herm ( \sum_{m=1}^K \sum_{n=1}^L \hat{\hv}_{jmn} \hat{\hv}_{jmn}^\herm - M z \Id_M )^{-1} \Lambdam_j \times \\
&~(\sum_{m=1}^K \sum_{n=1}^L \hat{\hv}_{jmn} \hat{\hv}_{jmn}^\herm - M z \Id_M )^{-1} \hat{\hv}_{jkp}\\
=\lim_{z \rightarrow 0} & ~{1\over M} \tilde{\hv}_{jkp}^\herm (\Bm_M-zI_M)^{-1}\Lambdam_j (\Bm_M-zI_M)^{-1} \tilde{\hv}_{jkp},
\end{align*}
where $\tilde{\hv}_{jmn} = \frac{1}{\sqrt{M}} \hat{\hv}_{jmn}\sim {\cal CN}(0, \frac{1}{M} \bar{\Rm}_{nn}^m = \Rm_{jmn} \Km_{mj}^{-1}$ $\Rm_{jmn} p_{mn})$ and
$\Bm_M = \sum_{m=1}^K \sum_{n=1}^L \tilde{\hv}_{jmn} \tilde{\hv}_{jmn}^\herm$. 

We will omit $j$ to shorten  notations. Below, we first obtain an approximation to the above quantity using tools from random matrix theory \cite{wagner2012large} for finite $z$ and then take the limit $z \rightarrow 0$.
Note that the cross covariance between vectors $\tilde{\hv}_{kp}$ and $\tilde{\hv}_{mn}$ is given by
{\small{
\begin{equation}
\EE[\tilde{\hv}_{kp} \tilde{\hv}_{mn}^\herm] = \left\{ \begin{array}{ll}
\frac{1}{M} \bar{\Rm}_{pn}^k = \frac{1}{M} \Rm_{jkp} \Km_{kj}^{-1} \Rm_{jkn} \sqrt{p_{kn} p_{kp}}, & k = m,\\
\zerov, & k \neq m.
\end{array} \right. \nonumber
\end{equation}}}
We define $m_M(z,-\alpha \Lambdam,\bar{\Rm}_{qn}) = \frac{1}{M} \trace{[\bar{\Rm}_{qn} ( \Bm_M - \alpha \Lambdam}$ ${ - z \Id_M )^{-1}]}$, where $\alpha$ is a positive scalar. For a finite $L$, and $M,K \longrightarrow \infty$, $M/K=\mbox{const}$,  we define a deterministic equivalent for $m_M(z,-\alpha \Lambdam,\bar{\Rm}_{qn})$, denoted by $f_{qn, -\alpha \Lambdam}^k$, by
\begin{equation} \label{eqn:det-eq-x}
m_M(z,-\alpha \Lambdam,\bar{\Rm}_{qn}) - f_{qn, -\alpha \Lambdam}^k \stackrel{M \longrightarrow \infty}{=} 0 \ \ {\rm a.s.}
\end{equation}
We further prove that
\begin{align} \label{eqn:fixed-pt}
f_{qn,-\alpha \Lambda}^k &= \frac{1}{M} \trace(\bar{\Rm}_{qn}^k ( \frac{1}{M} \sum_{k=1}^K \sum_{n=1}^L [\bar{\Rm}_{nn}^k -  \nonumber\\
& \sum_{p=1}^L c_{p,-\alpha \Lambdam}^{nk} \bar{\Rm}_{np}^k ] - \alpha \Lambdam - z \Id_M)^{-1} ), \nonumber\\
\cv_{-\alpha \Lambdam}^{nk} &= ( \Id_L + \Fm_{k,-\alpha \Lambdam}^T )^{-1} \fv_{n,-\alpha \Lambdam}^k,
\end{align}
with $\cv_{-\alpha \Lambdam}^{nk} = [c_{1,-\alpha\Lambdam}^{nk}\ \ldots\ c_{L,-\alpha \Lambdam}^{nk}]^T$, $\fv_{n,-\alpha \Lambdam}^k = [f_{1n,-\alpha \Lambdam}^k\ \ldots\ f_{Ln,-\alpha \Lambdam}^k]^T$ and $L\times L$ matrix  $[\Fm_{k,-\alpha \Lambdam}]_{nq} = f_{qn,-\alpha \Lambdam}^k$. Note that (\ref{eqn:fixed-pt}) is a fixed point equation and since $k = 1,\ldots,K,\ n,p = 1,\ldots,L$, we have $KL^2$ such equations. Hence we can find  $f_{qn,-\alpha\Lambdam}^k$'s as solutions of these equations.

We define $\Tm_{-\alpha\Lambdam} = \frac{1}{M} \sum_{k=1}^K \sum_{n=1}^L [\bar{\Rm}_{nn}^k - \sum_{p=1}^L c_{p,-\alpha \Lambdam}^{nk} \bar{\Rm}_{np}^k ]$ $ - \alpha \Lambdam - z \Id_M$, $\Tm = \Tm_{\zerov}$, $f_{qn}^k = f_{qn,\zerov}^k$, $c_p^{nk} = c_{p,\zerov}^{nk}$. Differentiating (\ref{eqn:det-eq-x}) and taking the value at $\alpha = 0$, we get
\begin{align} \label{eqn:diff-alpha}
\frac{d}{d \alpha}   \left[ m_M(z,-\alpha \Lambdam,\bar{\Rm}_{qn}^k) - \right.&\left. f_{qn, -\alpha \Lambdam}^k \right] \vrule_{\alpha = 0} \stackrel{M \longrightarrow \infty}{=} 0 \ \ {\rm a.s.} \nonumber\\
\Longrightarrow \frac{d}{d \alpha} \frac{1}{M} \trace\left[\bar{\Rm}_{qn}^k \left( \Bm_M - \right. \right. & \left. \left. \alpha \Lambdam - z \Id_M \right)^{-1}\right] \vrule_{\alpha = 0} \nonumber\\
&\stackrel{M \longrightarrow \infty}{=} \frac{d}{d \alpha} f_{qn, -\alpha \Lambdam}^k \vrule_{\alpha = 0} \nonumber\\
\Longrightarrow \frac{1}{M} \trace\left[ \bar{\Rm}_{qn}^k \left( \Bm_{M} - \right. \right. & \left. \left. z \Id_M \right)^{-1}  \Lambdam \left( \Bm_{M} - z \Id_M \right)^{-1} \right] \nonumber\\
&\stackrel{M \longrightarrow \infty}{=} \bar{f}_{qn,\Lambdam}^k,
\end{align}
where $\frac{d}{d \alpha} f_{qn, -\alpha \Lambdam}^k \vrule_{\alpha = 0} = \bar{f}_{qn,\Lambdam}^k$. We define vector $\fv_{n}^k = [f_{1n}^k\ \ldots\ f_{Ln}^k]^T$ and matrices $[{\Fm}_{k}]_{nq} = {f}_{qn}^k$, $[\bar{\Fm}_{k}]_{nq} = \bar{f}_{qn}^k$. 
We find  $\bar{f}_{qn,\Lambdam}^k$ by taking the derivative of both sides of (\ref{eqn:fixed-pt}) with respect to $\alpha$ and taking the limit $\alpha \rightarrow 0$. This gives
\begin{align} \label{eqn:fixed-pt-2}
\bar{f}_{qn,\Lambdam
}^k = &\frac{1}{M} \trace{\left( \bar{\Rm}_{qn}^k \Tm^{-1} \Lambdam \Tm^{-1} \right)} \nonumber\\
&+\sum_{m=1}^K \sum_{p=1}^L \left[ \uv_p^{m T} \bar{\fv}_{p,\Lambdam}^m - \uv_p^{m T} \bar{\Fm}_{m,\Lambdam}^T \wv_p^m \right],
\end{align}
where $\vv_p^m = [v_{p1}^m\  \ldots\ v_{pL}^m]^T$, $\wv_p^m = \left( \Id_L + \Fm_m^T \right)^{-1} \fv_p^m$ and $\uv_{p}^m = \left( \Id_L + \Fm_m \right)^{-1} \vv_p^m$ are such that
$v_{pl}^m = \frac{1}{M} \trace{\left(\bar{\Rm}_{qn}^k \Tm^{-1} \bar{\Rm}_{pl}^m \Tm^{-1} \right)}$.
Note that (\ref{eqn:fixed-pt-2}) defines $KL^2$ linear equations. Hence, using $f_{qn}^k$, we can find  $\bar{f}_{qn,\Lambdam}^k$.

One can see that the coefficients $f_{qn}^k, \bar{f}_{qn,\Lambdam}^k$ are functions of $z$. We define quantities  $\tilde{f}_{qn}^k=\lim_{z \rightarrow 0} -z f_{qn}^k$ and $\tilde{\bar{f}}_{qn,\Lambdam}^k=\lim_{z \rightarrow 0} z^2 \bar{f}_{qn,\Lambdam}^k $ and matrices $[\tilde{\Fm}_{k}]_{nq} = \tilde{f}_{qn}^k$,
$[\tilde{\bar{\Fm}}_{k,\Lambdam}]_{nq} = \tilde{\bar{f}}_{qn,\Lambdam}^k$. 
 To obtain $\tilde{f}_{qn}^k$, we multiply both sides of (\ref{eqn:fixed-pt}) by $-z$ and take the limit $\alpha,z \rightarrow 0$. Similarly, to obtain the quantity $\tilde{\bar{f}}_{qn,\Lambdam}^k$, we multiply both sides of (\ref{eqn:fixed-pt-2}) by $z^2$ and take the limit $z \rightarrow 0$.

Defining $\Bm_{k,M} = \Bm_M - \sum_{n=1}^L \tilde{\hv}_{kn} \tilde{\hv}_{kn}^\herm$, and making use of (\ref{eqn:diff-alpha}), we get that
\begin{align} \label{eqn:B_kM} 
& \tilde{\hv}_{kn}^\herm \left( \Bm_{k,M} - z \Id_M \right)^{-1} \Lambdam \left( \Bm_{k,M} - z \Id_M \right)^{-1} \tilde{\hv}_{kp} \nonumber\\
=& \trace \left[\tilde{\hv}_{kp} \tilde{\hv}_{kn}^\herm  \left( \Bm_{k,M} - z \Id_M \right)^{-1} \Lambdam \left( \Bm_{k,M} - z \Id_M \right)^{-1} \right] \nonumber\\
\stackrel{M \rightarrow \infty}{=}& \frac{1}{M} \trace{\left[ \bar{\Rm}_{pn}^k \left( \Bm_{k,M} - z \Id_M \right)^{-1} \Lambdam \left( \Bm_{k,M} - z \Id_M \right)^{-1} \right]} 
\nonumber\\
= & \bar{f}_{pn,\Lambdam}^k.
\end{align}
Let $\tilde{\Hm}_k=(\tilde{\hv}_{k1}\ \ldots\ \tilde{\hv}_{kL})$, $\Dm_k=(\Bm_{k,M}-z\Id_M)^{-1}$, $\Em_k=(\Id_L+\tilde{\Hm}_k^\herm\Dm_k\tilde{\Hm}_k)^{-1}$, and
$\Gm_k=\Dm_k\tilde{\Hm}_k \Em_k \tilde{\Hm}_k^\herm \Dm_k$. Now, using (\ref{eqn:B_kM}), after rigorous computations, we obtain 
\begin{align*}
& \frac{1}{M} \tilde{\hv}_{kn}^\herm \left( \Bm_{M} - z \Id_M \right)^{-1} \Lambdam \left( \Bm_{M} - z \Id_M \right)^{-1} \tilde{\hv}_{kp} \nonumber\\
=& \frac{1}{M} (\tilde{\hv}_{kn}^\herm \Dm_k\Lambdam\Dm_k\tilde{\hv}_{kp}-\tilde{\hv}_{kn}^\herm\Gm_k\Lambdam\Dm_k\tilde{\hv}_{kp} 
-\tilde{\hv}_{kn}^\herm\Dm_k\Lambdam\Gm_k\tilde{\hv}_{kp}\nonumber \\
&+\tilde{\hv}_{kn}^\herm \Gm_k\Lambdam\Gm_k \tilde{\hv}_{kp})  = \bar{f}_{pn,\Lambdam}^k -  
\bar{\fv}_{n,\Lambdam}^{k T} \left( \Id_L + \Fm_k \right)^{-1} \fv_{p}^{k *} \nonumber \\
 &- \fv_{n}^{k T} \left( \Id_L + \Fm_k \right)^{-1} \bar{\fv}_{p,\Lambdam}^{k *}   \nonumber \\
&+\fv_{n}^{k T} \left( \Id_L
+ \Fm_k \right)^{-1} \bar{\Fm}_{k,\Lambdam} \left( \Id_L + \Fm_k \right)^{-1} \fv_{p}^{k *}.
\end{align*}
As $z \rightarrow 0$, we can prove that the above quantity reduces to
\begin{align}
& \lim_{z \rightarrow 0} \frac{1}{M} \tilde{\hv}_{kn}^\herm \left[ \Bm_M - z \Id_M \right]^{-1} \Lambdam \left[ \Bm_M - z \Id_M \right]^{-1} \tilde{\hv}_{kp} \nonumber\\ &= \ev_n^T (\tilde{\Fm}_k)^{-1} \tilde{\bar{\Fm}}_{k,\Lambdam} (\tilde{\Fm}_k)^{-1} \ev_p = \ev_n^T \Gammam_{k} \ev_p, \label{eqn:approx-val2}
\end{align}
where  $\ev_n$ is the $n^{\rm th}$ column of $\Id_L$, and $\Gammam_{k} = (\tilde{\Fm}_k)^{-1} \tilde{\bar{\Fm}}_{k,\Lambdam} (\tilde{\Fm}_k)^{-1}$.   Note that $\Gammam_k$ is  $\Gammam_{jk}$ in notations of  (\ref{eqn:approx-val3}) (we dropped the index $j$ in the beginning of these derivations). Note also that $\Gammam_{jk}$ depends only on slow fading components.


\begin{thebibliography}{1}
	\providecommand{\url}[1]{#1}
	\csname url@samestyle\endcsname
	\providecommand{\newblock}{\relax}
	\providecommand{\bibinfo}[2]{#2}
	\providecommand{\BIBentrySTDinterwordspacing}{\spaceskip=0pt\relax}
	\providecommand{\BIBentryALTinterwordstretchfactor}{4}
	\providecommand{\BIBentryALTinterwordspacing}{\spaceskip=\fontdimen2\font plus
		\BIBentryALTinterwordstretchfactor\fontdimen3\font minus
		\fontdimen4\font\relax}
	\providecommand{\BIBforeignlanguage}[2]{{%
			\expandafter\ifx\csname l@#1\endcsname\relax
			\typeout{** WARNING: IEEEtran.bst: No hyphenation pattern has been}%
			\typeout{** loaded for the language `#1'. Using the pattern for}%
			\typeout{** the default language instead.}%
			\else
			\language=\csname l@#1\endcsname
			\fi
			#2}}
	\providecommand{\BIBdecl}{\relax}
	\BIBdecl
	
	\bibitem{marzetta2010noncooperative}
	T.~L. Marzetta, ``Noncooperative cellular wireless with unlimited numbers of
	base station antennas,'' \emph{Wireless Communications, IEEE Transactions
		on}, vol.~9, no.~11, pp. 3590--3600, 2010.
	
	\bibitem{ashikhmin2012pilot}
	A.~Ashikhmin and T.~Marzetta, ``Pilot contamination precoding in multi-cell
	large scale antenna systems,'' pp. 1137--1141, 2012.
	
	\bibitem{adhikary}
	A.~Adhikary, A.~Ashikhmin, and T.~L. Marzetta, ``Uplink interference reduction
	in large-scale antenna systems,'' \emph{Communications, IEEE Transactions
		on}, vol.~16, pp. 2194 -- 2206, 2017.
	
	\bibitem{ashikhmin}
	A.~Ashikhmin, L.~Li, and T.~L. Marzetta, ``Interference reduction in large
	scale antenna systems,'' \emph{Information Theory, IEEE Transactions on}, to
	appear, 2018.
	
	\bibitem{hoydis}
	E.~Björnson, J.~Hoydis, and L.~Sanguinetti, ``Massive mimo has unlimited
	capacity,'' \emph{Wireless Communications, IEEE Transactions on}, vol.~17,
	no.~1, pp. 1536--1276, 2017.
	
	\bibitem{shiu2000fading}
	D.-S. Shiu, G.~J. Foschini, M.~J. Gans, and J.~M. Kahn, ``Fading correlation
	and its effect on the capacity of multielement antenna systems,''
	\emph{Communications, IEEE Transactions on}, vol.~48, no.~3, pp. 502--513,
	2000.
	
	\bibitem{wagner2012large}
	S.~Wagner, R.~Couillet, M.~Debbah, and D.~T. Slock, ``Large system analysis of
	linear precoding in correlated miso broadcast channels under limited
	feedback,'' \emph{IEEE transactions on information theory}, vol.~58, no.~7,
	pp. 4509--4537, 2012.
	
\end{thebibliography}


\end{document}